\documentclass[fleqn,usenatbib]{mnras}

\usepackage{newtxtext,newtxmath}

\usepackage[T1]{fontenc}

\DeclareRobustCommand{\VAN}[3]{#2}
\let\VANthebibliography\thebibliography
\def\thebibliography{\DeclareRobustCommand{\VAN}[3]{##3}\VANthebibliography}

\usepackage{graphicx}	
\usepackage{amsmath}	

\usepackage{acronym}

\newcommand{\msun}{\textnormal{M}_\odot}
\newcommand{\rsun}{\textnormal{R}_\odot}
\newcommand{\lsun}{\textnormal{L}_\odot}

\acrodef{ccsn}[CCSN]{core-collapse supernova}
\acrodefplural{ccsn}[CCSNe]{core-collapse supernovae}
\acrodef{sn}[SN]{supernova}
\acrodefplural{sn}[SNe]{supernovae}
\acrodef{sesn}[SESN]{stripped-envelope supernova}
\acrodefplural{sesn}[SESNe]{stripped-envelope supernovae}
\acrodef{ns}[NS]{neutron star}
\acrodefplural{ns}[NSs]{neutron stars}
\acrodef{bh}[BH]{black hole}
\acrodefplural{bh}[BHs]{black holes}
\acrodef{eci}[ECI]{ejecta--companion interaction}
\acrodef{ms}[MS]{main-sequence}
\acrodef{zams}[ZAMS]{zero-age main-sequence}
\acrodef{hrd}[HRD]{Hertzsprung--Russell diagram}
\acrodef{mt}[MT]{mass transfer}
\acrodef{csm}[CSM]{circumstellar material}

\title[Modelling post-SN inflated companions]{Constraining mass-transfer and common-envelope physics with post-supernova companion monitoring}

\author[R. Hirai]{
Ryosuke Hirai$^{1,2}$\thanks{E-mail: ryosuke.hirai@monash.edu}
\\
$^{1}$ School of Physics and Astronomy, Monash University, Clayton, Victoria 3800, Australia\\
$^{2}$ OzGrav: The ARC Centre of Excellence for Gravitational Wave Discovery, Australia
}

\date{Accepted XXX. Received YYY; in original form ZZZ}

\pubyear{2022}

\begin{document}
\label{firstpage}
\pagerange{\pageref{firstpage}--\pageref{lastpage}}
\maketitle

\begin{abstract}
We present an analytical model that describes the response of companion stars after being impacted by a supernova in a close binary system. This model captures key properties of the luminosity evolution obtained from 1D stellar evolution calculations fairly well: a high-luminosity plateau phase and a decaying tail phase. It can be used to constrain the pre-supernova binary properties from the observed photometry of the companion star several years after the explosion in a relatively simple manner. The derived binary parameters are useful in constraining the evolutionary scenario for the progenitors and the physics of binary interactions. We apply our model to some known stripped-envelope supernova companions (SN1993J, SN2001ig, SN2006jc, SN2011dh, SN2013ge). Combined with other observational constraints such as the pre-supernova progenitor photometry, we find that SN1993J and SN2011dh likely had relatively massive companions on wide orbits, while SN2006jc may have had a relatively low-mass companion on a tight orbit. This trend suggests that type IIb supernova progenitors evolved from stable mass transfer channels and type Ibc progenitors may have formed from common-envelope channels. The constraints on orbital separation helps us probe the highly uncertain common-envelope physics for massive stars, especially with multiple epochs of companion observations. We also highlight possible limitations of our model due to the assumptions made in the underlying 1D models.
\end{abstract}

\begin{keywords}
supernovae: general -- binaries: general -- supernovae: individual: SN1993J, SN2001ig, SN2006jc, SN2011dh, SN2013ge
\end{keywords}



\section{Introduction}

The vast majority of massive stars are born as members of binary or higher order multiple systems \citep[]{san12,moe17,off22}. Among them, the systems with relatively tight orbits are particularly interesting, as they can experience various forms of binary interactions, exchanging and/or ejecting matter from the system. This is considered to be one of the main reasons why there is a wide diversity of observational properties for \acp{ccsn} \citep[e.g.][]{pod92}.

\begin{figure}
 \centering
 \includegraphics[width=\linewidth]{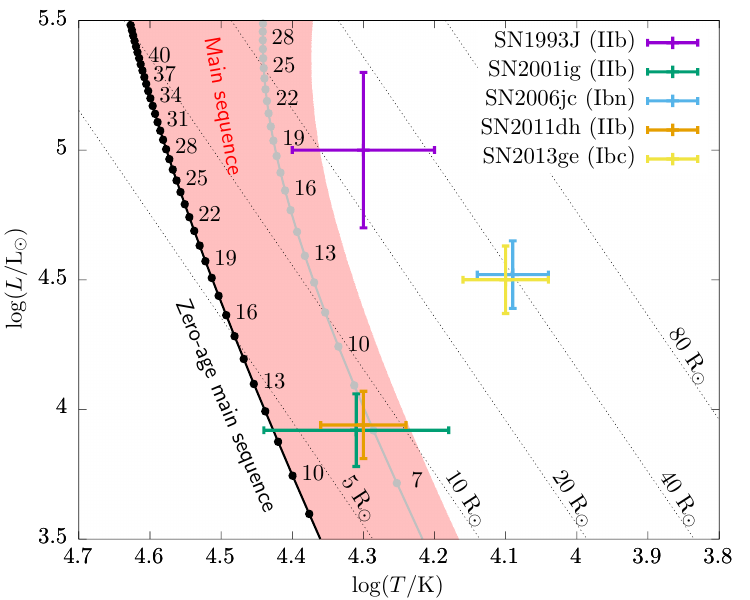}
 \caption{Location of detected \ac{sesn} companions on the \ac{hrd}, based on Table~\ref{tab:photometry}. The black curve indicates the zero-age main sequence whereas the grey curve indicates the terminal-age main sequence, and the numbers are masses in solar units. The pink shaded region shows the area that can be covered by main-sequence stars. For SN2013ge, we added error bars similar to SN2006jc for visibility.\label{fig:companions}}
\end{figure}

It is expected that many \acp{ccsn} are occurring with the presence of a closeby companion star, especially for \acp{sesn}, where their progenitors have lost most or all of their hydrogen-rich envelopes prior to explosion \citep[e.g.][]{zap17,zap19}. Indeed, some companion stars have been discovered in the location of \acp{sesn}, several years after their explosions. There are five candidates detected so far: SN1993J \citep[]{mau04,fox14}, SN2001ig \citep[]{ryd18}, SN2006jc \citep[]{mau16,sun20}, SN2011dh \citep[]{mau19} and SN2013ge \citep[]{fox22}. We summarize the properties of the explosion and surviving companion in Table~\ref{tab:photometry} and plot them on the \ac{hrd} in Figure~\ref{fig:companions}. While two of the companions (SN2001ig, SN2011dh) are consistent with being a main-sequence star (pink shaded region), the other three are located within the Hertzsprung gap. This is in tension with what is typically expected, as stars only spend a very short fraction of their lives in that phase \citep[e.g.][]{zap17}. For some of these \acp{sn} (SN1993J, SN2011dh) the pre-\ac{sn} progenitor was also detected \citep[]{ald94,mau11,van11}, providing critical information for understanding the evolution of the binary up to the explosion \citep[e.g.][]{pod93,mau04,ben13}. However, these models are typically constructed under an implicit assumption that the companion is unaffected by the \ac{sn}.

In tight binaries, the \ac{sn} ejecta can collide with the companion and alter its properties. Hydrodynamical simulations show that direct mass-stripping is typically negligible when the companion is a main-sequence star \citep[$\lesssim1~\%$;][]{RH15,liu15,rim16,RH18,che23}. Instead, the main effect of the \ac{sn} is to heat the star through shocks, temporarily driving it out of thermal equilibrium. In response, the star can inflate and appear much cooler and more luminous than its original state \citep[]{RH15,rim16,RH18,oga21,che23}. This inflated state can only be maintained for a short while, and will quickly shrink back after a couple of years to decades. In \citet{oga21} we characterized this response with two parameters (the maximum luminosity and inflated duration), and found that they are strongly correlated with the mass of the companion and pre-SN orbital separation. Therefore, by observing the response of companion stars after \acp{sn}, we may be able to constrain the pre-SN orbital properties that are otherwise inaccessible.

Any empirical constraints on the pre-\ac{sn} orbital separation is extremely valuable for understanding binary evolution. For example, one of the most important yet poorly understood processes in binary evolution is common-envelope evolution \citep[]{iva13,iva20}. In this phase, the primary star evolves and expands to engulf the companion star, leading to a rapid inspiral of the orbit and efficient mass ejection. It is recently being heavily studied in the massive star regime \citep[e.g.][]{fra19,law20,lau22a,lau22b,mor22,RH22}, as it is considered to be responsible for explaining the origin of tight binaries involving compact objects such as X-ray binaries, \ac{sesn} progenitors, gravitational wave sources, etc. Predicting the post-common-envelope orbital separation is an area of active research \citep[]{RH22,dis23}, and thus any observational constraint will be useful for distinguishing among these models.

The more stable form of mass transfer such as Roche lobe overflow is also a relatively uncertain process. Even in a steady mass transfer situation, not all the mass is retained by the accretor and part of the mass can leave the system, carrying away angular momentum. Depending on the mass retention rate and the angular momentum loss, the post-mass-transfer orbit and companion mass can vary significantly. If \ac{sesn} progenitors were formed through stable mass transfer, observational constraints on pre-\ac{sn} separations and the companion mass will help us understand these processes better.

In this paper, we build upon our previous work \citep[]{RH18,oga21} to construct a more complete model for the post-\ac{sn} companion response. The key input parameters for the model are: the companion mass, companion radius, explosion energy and orbital separation. Using the properties of the observed companions to \ac{sesn} progenitors, we place constraints on these input parameters. By combining this information with other observational clues, we are able to infer the history of the binary leading up to the \ac{sn}.

This paper is structured as follows. We first briefly review our previous work and explain our new extended analytical model in Section~\ref{sec:method}. Results are presented in Section~\ref{sec:results} and we discuss the implications in Section~\ref{sec:discussion}. We provide a summary and conclude our work in Section~\ref{sec:summary}.

\begin{table*}
 \begin{center}
  \caption{Summary of the supernova properties and companion photometry.\label{tab:photometry}}
  \begin{tabular}{ccccccc}
   \hline
   Name & SN type & $E_\mathrm{exp}$ (B) & $\log(L/\lsun)$ & $\log(T_\mathrm{eff}/\mathrm{K})$ & Epoch (yr) & Reference \\\hline
   SN1993J  & IIb & $1$--$1.3$ & $5\pm0.3$ & $4.3\pm0.1$ & 18.9 & \citet{you95,fox14}\\
   SN2001ig & IIb & $-$ & $3.92\pm0.14$ & $4.31\pm0.13$ & 14.4 & \citet{ryd18}\\
   SN2006jc & Ibn & $\sim10$& $4.52\pm0.13$ & $4.09\pm0.05$ & 10.4 & \citet{tom08,sun20}\\
   SN2011dh & IIb & $0.6$--$1$& $3.94\pm0.13$ & $4.30\pm0.06$ &  6.3 & \citet{ber12,mau19}\\
   SN2013ge & Ibc & $1$--$2$ & $\sim4.5$ & $\sim4.1$ & 7.0 & \citet{dro16,fox22}\\
   \hline
  \end{tabular}
 \end{center}
 
\end{table*}

\section{Method}\label{sec:method}
\subsection{Review of ejecta-companion interaction models}

In \citet{RH18} and \citet{oga21}, we discovered that when the companion is a main-sequence star, it can become temporarily inflated and overluminous after being impacted by the \ac{sn} ejecta of the primary star. According to our 2D hydrodynamical simulations \citep[]{RH18}, the ejecta deposits energy into the companion through shocks, inversely proportional to the mass from the surface\footnote{\citet{che23} notes that this distribution may change for the most strongly impacted cases.}. There is a tiny amount of mass that is stripped off or ablated away, but that is usually negligible ($\lesssim1~\%$). The excess specific energy distribution in the remaining star takes the form
\begin{equation}
 \Delta\epsilon(m)=\frac{E_\mathrm{heat}}{m_\mathrm{h}[1+\ln{(M_2/m_\mathrm{h})]}}\times \begin{cases}									  1, & \text{if $m\leq m_\mathrm{h}$}.\\
											  m_\mathrm{h}/m, & \text{if $m>m_\mathrm{h}$}.
											 \end{cases}
\label{eq:Eexcess}
\end{equation}
where $m$ is the mass from the surface, $M_2$ is the total companion mass, $E_\mathrm{heat}$ is the energy injected into the star, and $m_\mathrm{h}$ is a parameter that describes the efficiently heated mass. Due to the virial theorem, half of the injected energy is used for expansion and therefore only half of $E_\mathrm{heat}$ is dissipated as heat. The injected energy can be computed as $E_\mathrm{heat}=pE_\mathrm{exp}\tilde{\Omega}$, where $p$ is the energy deposition efficiency, $E_\mathrm{exp}$ is the explosion energy  and $\tilde{\Omega}$ is the fractional solid angle subtended by the companion described as
\begin{equation}
 \tilde{\Omega}=\frac{1-\sqrt{1-(R_2/a)^2}}{2}\sim \frac{R_2^2}{4a^2}.
\end{equation}
 Here $R_2$ is the companion radius and $a$ is the orbital separation. The value of the energy deposition efficiency ranges around $p\sim0.08$--$0.12$, depending on how much the star gets compressed by the \ac{sn} ejecta (see discussion in Section~4.3 of \citealt{RH18}). Similarly, we found that $m_\mathrm{h}$ was related to the ejecta mass $M_\mathrm{ej}$ through $m_\mathrm{h}=M_\mathrm{ej}\tilde{\Omega}/2$.

Most of the excess energy is located around the surface, expanding a small amount of the surface layers to extremely large radii. Typically in our numerical experiments using the 1D stellar evolution code MESA \citep[]{pax11,pax13,pax15,pax18}, the star stays at a fixed luminosity and radius for a few years to decades, before abruptly contracting back to close to its original size. We therefore characterized the response with two parameters: the maximum luminosity $L_\mathrm{max}$ and inflated duration $\tau_\mathrm{inf}$. We found that the maximum luminosity is almost only dependent on the companion mass $M_2$ through a relation
\begin{equation}
 L_\mathrm{max}=\frac{4\pi GM_2c}{\kappa_\mathrm{fit}},\label{eq:lmax}
\end{equation}
where $G$ is the gravitational constant, and $c$ the speed of light. This functional form is motivated by the fact that the stellar luminosity is restricted by the Eddington luminosity, except the relevant opacity is not trivial. In our 1D stellar models, the luminosity was limited by the local Eddington luminosity at the base of the near-surface convective layer, where the relevant opacity was somewhere at the edge of the so-called iron bump (see discussion in Section 4.1 of \citealt{oga21}). We empirically set this opacity parameter $\kappa_\mathrm{fit}$ as
\begin{equation}
 \kappa_\mathrm{fit}=\kappa_0\left(1-b\frac{M_2}{\msun}\right),
\end{equation}
where the coefficients were $\kappa_0=1.24~\mathrm{cm}^2~\mathrm{g}^{-1}$ and $b=0.02$ based on fits to the maximum luminosities obtained from a large set of MESA models \citep[]{oga21}. The inflated state ceases once the star radiates away a fraction $\alpha$ of the injected energy $E_\mathrm{heat}$. Thus the inflated duration can be approximated as
\begin{equation}
 \tau_\mathrm{inf}=\alpha\frac{E_\mathrm{heat}}{L_\mathrm{max}}.\label{eq:tau_inf_ogata}
\end{equation}
We found that most of our models can be well approximated with $\alpha\sim0.18$.

\begin{figure}
 \centering
 \includegraphics[width=\linewidth]{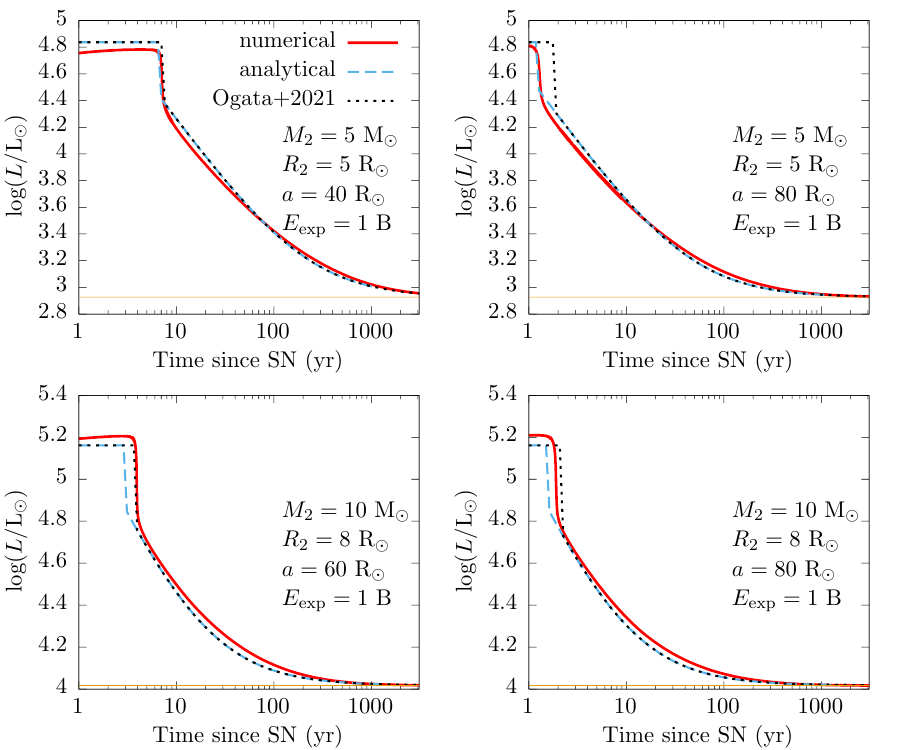}
 \caption{Light curves of the companion response to SN heating. Each panel shows results for different binary and explosion parameters. Red solid curves are from simulations with MESA and the black dashed curves are from our analytical model. The orange lines mark the luminosity of the companion before the \ac{sn} impact. \label{fig:LCfits}}
\end{figure}

We show some examples of the luminosity evolution of the companion in Figure~\ref{fig:LCfits} (solid red curves). The numerical models were computed in the same way as in our previous work \citep{RH18,oga21,oga21z}. We choose an injection efficiency of $p=0.08$ and all other parameters are described in each panel. The star is initially driven out of thermal equilibrium, reaching a luminosity close to the limit allowed by a star of that mass (Eq.~(\ref{eq:lmax})). As the star abruptly shrinks in size after $\tau_\mathrm{inf}$, the luminosity also drops abruptly by a factor of 2--3 (0.4--0.5 dex). However, it is still quite overluminous compared to its pre-heated state, as some of the energy deposited in the deeper layers diffuse out later on. The luminosity declines over a few hundred years until it regains thermal equilibrium.

\subsection{Analytical model}

Here we construct an analytical model to describe the full light curve of the companion including this tail part that we previously did not model. Following our previous work, we assume that no mass was stripped off and the heating of the companion by the SN is instantaneous. A fraction of the \ac{sn} energy is deposited into the star with a distribution following Eq.~(\ref{eq:Eexcess}). The star then tries to get rid of this excess heat by transporting it outwards. We assume a very simple picture where the excess energy is transported outwards in mass coordinate with some timescale $\tau_\mathrm{tr}$ that is universal over the entire star. It can be considered as carrying the excess energy on a conveyor belt, where the energy that reaches the end of the conveyor belt (= stellar surface) is emitted as radiation. The conveyor belt completes one cycle over the time $\tau_\mathrm{tr}$. Under this assumption, the surface excess luminosity at a given time can be described as 
\begin{equation}
 L_\mathrm{ex}(t)=\frac{\Delta\epsilon (M_2t/\tau_\mathrm{tr})M_2}{2\tau_\mathrm{tr}}.\label{eq:L_ex}
\end{equation}
The term $\Delta\epsilon (M_2t/\tau_\mathrm{tr})$ indicates which part of the initial excess energy distribution has reached the surface. This specific energy is multiplied by $M_2$ and divided by the transport timescale $\tau_\mathrm{tr}$ to obtain the rate at which the excess energy escapes the star.
Note that the factor two in the denominator comes from an assumption that half of the luminosity is used as thermodynamic work to maintain the inflated stellar structure.
However, the star does not radiate at luminosities higher than the maximum allowed value given in Eq.~(\ref{eq:lmax}) in our MESA models. Hence the luminosity is initially capped by $L_\mathrm{max}$ until the total radiated energy exceeds the integrated energy excess from the surface down to $m=M_2t/\tau_\mathrm{tr}$. It is analogous to having a ``waiting list'' for the excess heat to escape out, and the flux remains constant until the waiting list is exhausted. Once the waiting list is exhausted, the luminosity is determined by the amount of excess heat at the surface, as expressed in Eq.~(\ref{eq:L_ex}). The full light curve can thus be expressed as
\begin{equation}
 L_2(t,M_2,E_\mathrm{heat})=\begin{cases}
		       L_\mathrm{max}(M_2), & \text{if $t\leq\tau_\mathrm{inf}$}.\\
		       \dfrac{E_\mathrm{heat}}{2[1+\ln{(M_2/m_\mathrm{h})}]}\cdot \dfrac{1}{t}+L_\mathrm{org}, & \text{if $t>\tau_\mathrm{inf}$}.
		      \end{cases}
\label{eq:fullLC}
\end{equation}
where $L_\mathrm{org}=L_\mathrm{org}(M_2)$ is the luminosity of the companion prior to being heated.

The only model parameter that needs to be determined is the transport timescale $\tau_\mathrm{tr}$. Interestingly, it does not appear in Eq.~(\ref{eq:fullLC}) except implicitly in the definition of $\tau_\mathrm{inf}$. The $\tau_\mathrm{tr}$-dependence cancels out due to the heat excess distribution being proportional to $\Delta\epsilon\propto 1/m$ in Eq.~(\ref{eq:Eexcess}). This makes our light curve model rather robust to model assumptions at least for the tail part. We roughly estimate $\tau_\mathrm{tr}$ as
\begin{equation}
 \tau_\mathrm{tr}\sim\frac{GM_2^2}{2R_2L_\mathrm{max}(M_2)},\label{eq:Ttr_estimate}
\end{equation}
which is similar to the thermal timescale of the star, except evaluated with the maximum luminosity.
The inflated timescale can then be obtained by finding the larger root of the following equation for $\tau_\mathrm{inf}$
\begin{equation}
 (L_\mathrm{max}-L_\mathrm{org})\tau_\mathrm{inf}=\frac{E_\mathrm{heat}}{2[1+\ln{(M_2/m_\mathrm{h})}]}\left(1+\ln{\frac{M_2\tau_\mathrm{inf}}{m_\mathrm{h}\tau_\mathrm{tr}}}\right),\label{eq:Tinf_estimate}
\end{equation}
when 
\begin{equation}
 L_\mathrm{max}>\frac{E_\mathrm{heat}M_2}{2[1+\ln{(M_2/m_\mathrm{h})}]\tau_\mathrm{tr}m_\mathrm{h}}+L_\mathrm{org}.\label{eq:tau_inf_criterion}
\end{equation}
If Eq.~(\ref{eq:tau_inf_criterion}) is not satisfied, the star never approaches $L_2=L_\mathrm{max}$ so $\tau_\mathrm{inf}=0$. Note that this analytical framework is designed to mimic the results of our MESA models, so it is bound by the same uncertainties that the underlying 1D models have. The degree of inflation and $\tau_\mathrm{inf}$ can contain large uncertainties as explained in Section~\ref{sec:caveats}.

For the companion radius $R_2$ and luminosity $L_\mathrm{org}$, we use the analytical fits by \citet{hur00} for main-sequence stars, assuming an age $t_\mathrm{age}\sim10~$Myr, which roughly corresponds to the lifetime of a $20~\msun$ star. This assumes that the companion is coeval with the \ac{sn} progenitor and has not been rejuvenated by mass accretion. In reality, the companion may have accreted a substantial amount of mass depending on the preceding evolution, and our stellar radii and luminosity estimates may be incorrect. However, given the uncertainty in the progenitor mass (and thus companion age) and explosion energy, we ignore the errors on our companion property estimates.

\section{Results}\label{sec:results}

 We compare this analytic light curve model (dashed curves) against the numerical models (solid curves) in Figure~\ref{fig:LCfits}. Most remarkably, the tail part of the light curves display an almost perfect match for all the models shown. This demonstrates the validity of our model despite having many simplifying assumptions. For the plateau phase, we have recalibrated our fitting coefficients for $L_\mathrm{max}$ to $\kappa_0=1~\mathrm{cm}^2~\mathrm{g}^{-1}$ and $b=0.01$ (it was $\kappa_0=1.24~\mathrm{cm}^2~\mathrm{g}^{-1}$ and $b=0.02$ in \citealt{oga21}). With this recalibration, the analytical models show good agreement with the numerical models for $L_\mathrm{max}$ within the parameter space we explored. As for the duration of the plateau phase, our estimates based on Eqs.~(\ref{eq:Ttr_estimate})--(\ref{eq:Tinf_estimate}) seem to explain the numerical models relatively well although in many cases it slightly underestimates. As reference, we overplot some models estimating the inflated timescale with the empirical fit as in Eq.~(\ref{eq:tau_inf_ogata}) (dotted curves). In both cases, the light curve in the tail phase are treated identically.

\begin{figure}
 \centering
 \includegraphics[width=\linewidth]{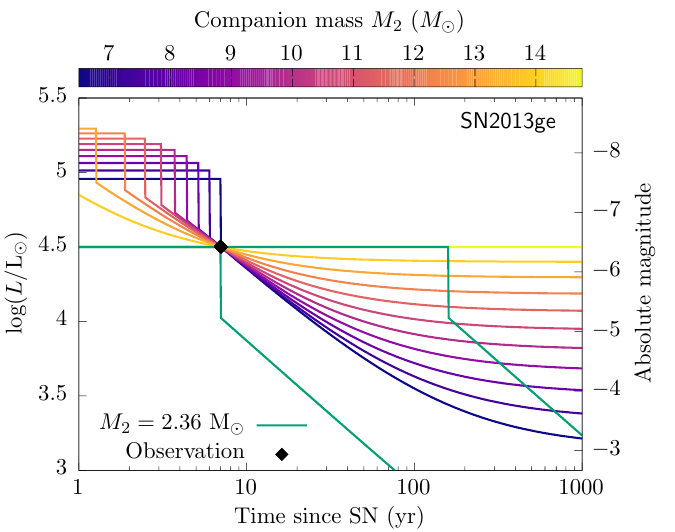}
 \caption{Companion response light curves that satisfy the current observed companion photometric constraints of SN2013ge (Black diamond). We assumed $E_\mathrm{exp}=1.5~$B and $M_2=4.5~\msun$. The green curves are example light curves fitted to the observation in the plateau phase with a companion mass of $M_2=2.36~\msun$ and two different orbital separations. The right (left) curve is computed with the minimum (maximum) orbital separation (see text for details). All other curves are computed by fitting the observation in the tail phase. \label{fig:sn2013ge_LC}}
\end{figure}

In Figure~\ref{fig:sn2013ge_LC}, we plot a set of analytical light curves that satisfy the observational constraint for the companion of SN2013ge \citep[]{fox22}. We assumed $E_\mathrm{exp}=1.5~$B, based on the observationally inferred explosion energy (see Table~\ref{tab:photometry}). There are two different ways to fit the single observed point, depending on which part of the light curve it is fitted in: the plateau phase or the tail phase.

The green curves were computed assuming the observed companion is in the plateau phase. Since the luminosity of the plateau phase only depends on the companion mass, the luminosity provides a unique constraint on the companion mass as $M_2=2.36~\msun$. The pre-SN orbital separation and explosion energy determines the duration of the plateau phase. Two light curves computed with the minimum and maximum possible orbital separations are shown, but any curve in between are also possible. The physically possible minimum separation is where the companion radius equals its Roche lobe radius, whereas the maximum separation is when the plateau phase duration is equal to the observed epoch for the companion star. To compute the Roche lobe radius, we assumed a progenitor mass of $M_1=4.5~\msun$, which produces an ejecta mass consistent with the typical observed value for \acp{sesn} \citep[$M_\mathrm{ej}\sim3~\msun$;][]{lym16,tad18,pre19}. At the minimum separation, the inflated duration can be as long as $\sim150~$yr. 

All other curves were fitted in the tail phase. A different combination of $a$ and $M_2$ is required for each curve to fit the observed point. The minimum possible mass in this regime is $M_2\sim6.5~\msun$, and can reach as high as $M_2\sim14.8~\msun$. Depending on the mass, the light curve will have a different decline rate and thus predict different luminosities 10~yr from now. With another epoch of observations, we will be able to constrain the true companion mass and pre-SN orbital separation at the same time. In Appendix~\ref{app:otherLC}, we show similar plots fitted to the observed companions for the other \acp{sn} listed in Table~\ref{tab:photometry}.

\begin{figure}
 \centering
 \includegraphics[width=\linewidth]{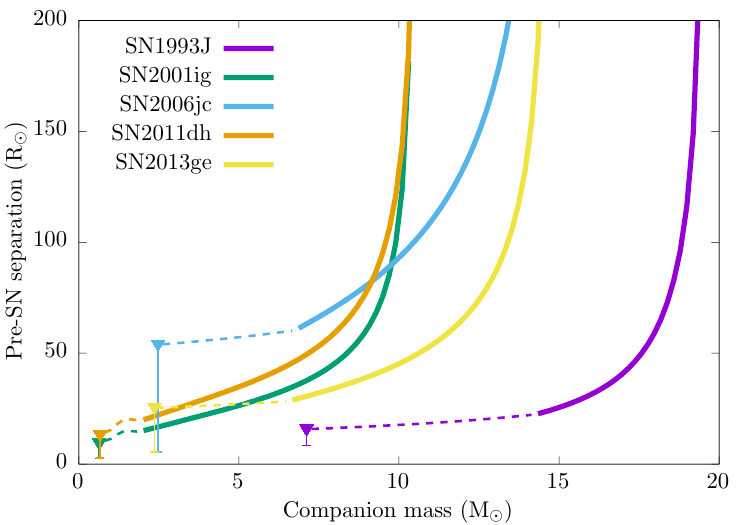}
 \caption{Constraints on the pre-\ac{sn} orbital separations. Here we assumed the progenitor had a lifetime of 8.7~Myr. Solid parts of the curves are based on fits to the tail phase. The vertical bars are based on fits to the plateau phase. The dashed parts of the curves are based on fits to the jump between the plateau and tail phases.\label{fig:sep_constraints}}
\end{figure}

Figure~\ref{fig:sep_constraints} shows the constraints on the allowed combinations of $a$ and $M_2$ based on our light curve fits to observed companions. The vertical bars with triangles denote the allowed range for pre-SN separations when fitted to the plateau phase. Solid parts of the curves indicate the constraints when fitted to the tail phase, whereas the dashed parts are fitted exactly at the location of the jump. All curves have similar shapes extending from the bottom left to top right. Basically, the current photometry can be explained by either a low-mass companion in a tight pre-SN orbit or a high-mass companion in a wide orbit. At the highest masses, the equilibrium luminosity $L_\mathrm{org}$ approaches the observed luminosity. At this limit, we can only place lower limits on the pre-SN separation based on the fact that the companion is essentially unheated \citep[]{riz23}.

\section{Discussion}\label{sec:discussion}

\subsection{Additional constraints}
In the previous section, we have only used a single post-SN companion photometric constraint to infer the pre-SN binary properties. For some of the \acp{sn} we have additional observational constraints that help us partially break the degeneracy.

\subsubsection{Pre-SN photometry}\label{sec:pre-explosion_image}

SN1993J and SN2011dh have pre-SN progenitor detections in addition to the companion. The size of stars inferred from the photometry can be used to rule out part of the parameter space, as the progenitor should have been contained within its Roche lobe prior to explosion. Roughly speaking, the separation should be $\gtrsim3$ times larger than the stellar radius in order to fit in a star within its Roche lobe, although the exact value depends on the mass ratio \citep[]{egg83}. Here we show that even the loosest constraint of $a>R_2$ is sufficient to rule out significant amounts of parameter space.

 The pre-explosion photometry of SN1993J is consistent with a K-supergiant with $\log{(L/\lsun)}=5.1\pm0.3$ and $\log{(T/\mathrm{K})}=3.63\pm0.05$ \citep[]{mau04}. This corresponds to a radius of $\sim650~\rsun$, meaning the pre-\ac{sn} separation had to be at least $a>650~\rsun$. At such wide separations, the \ac{sn} ejecta has little impact on the companion, and should look like a regular star. According to Figure~\ref{fig:sep_constraints}, the companion mass should be $M_2\sim19$--$20~\msun$ if the pre-SN separation was $>650~\rsun$. 

Similarly, the progenitor of SN2011dh is consistent with being an F-supergiant with $\log{(L/\lsun)}=4.92\pm0.20$ and $T=6000\pm260$~K \citep[]{mau11,van11}, with a corresponding radius of $R_2\sim270~\rsun$. This again is a sufficiently large value that the \ac{sn} ejecta would have little impact on the companion radius. The inferred companion mass would then be $\sim10~\msun$, based on Figure~\ref{fig:sep_constraints}.

\subsubsection{Multiple epochs of companion detections}\label{sec:multiple-epoch}

Our analytical constraints on the pre-\ac{sn} binary properties were placed based on one epoch of post-\ac{sn} observations (Table~\ref{tab:photometry}). Some of the \acp{sn} already have a second epoch of observations, which can be used to pin down the pre-\ac{sn} binary parameters to a single combination.

SN1993J is one of the best observed \acp{sn} aside from SN1987A, and has multiple late-time observations. The putative hot companion was first detected in 2003, with near-UV spectra suggesting the existence of a B-supergiant \citep[]{mau04}. It was later followed up in 2004--2006 \citep[]{mau09} and 2012 \citep[]{fox14}, confirming the hot B star contribution. Although the \ac{sn} and companion have not been fully disentangled, the B star contribution seems to be constant over the multiple epochs. This supports the scenario above that the companion is a relatively massive star ($M_2\sim19~\msun$) on a wide orbit such that it does not get strongly impacted by the \ac{sn}. 

SN2006jc is another object with multiple late-time observations. The first observation of the companion was made $\sim3.6$~yr after the explosion \citep[]{mau16}, whereas the second observation was $\sim10.4$~yr after the explosion \citep[]{sun20}. Different filter combinations were used in each epoch making it difficult to provide a direct comparison, but generally the fluxes were consistent with being the same object. If this is true, only the two ends of the solution space remain: a low-mass companion ($M_2=2.47~\msun$) on a tight orbit ($a\lesssim50~\rsun$) or a high-mass companion ($M_2\sim15~\msun$) on a wide orbit ($a\gtrsim200~\rsun$). Both solutions are viable, but the fact that the companion has a temperature off the main sequence (Fig~\ref{fig:companions}) supports the case that the star was somehow affected by the \ac{sn}. While it is not impossible to have a companion star that happens to be in the Hertzsprung gap, it is quite rare and requires a rather fine-tuned initial mass ratio. Therefore, we argue that the companion is more likely to have been a heated low-mass star rather than an unscathed high-mass star. Having an evolved companion also makes it prone to mass stripping by the \ac{sn} \citep[]{RH14,RH20}, and thus using our current analytical model is misleading.

\subsection{Caveats}\label{sec:caveats}

In our analytical model, we only modelled the luminosity evolution and ignored the radius/temperature response. Our previous 1D simulations show that the radius can be greatly inflated during the plateau phase and quickly contracts after the luminosity drop. Therefore, the star only stays in the Hertzsprung gap region for a relatively short time. It may thus be difficult to explain the exact photometry of observed companions as displayed in Figure~\ref{fig:companions}. In \citet{oga21}, we proposed that interactions with the new-born neutron star may be able to keep a star in the Hertzsprung gap region if the binary can survive. The current radius may then be a measure of the current orbit periapsis or the Roche lobe radius. 

The stellar radii of these temporarily out-of-equilibrium stars are also very sensitive to the outer boundary conditions. Since the luminosity of the surface layers of the heated star is close to the Eddington limit, the 1D models computed based on hydrostatic equilibrium are not completely reliable. The high luminosity may drive non-spherical effects, altering the effective opacity distribution as we discuss later. Alternatively, the high luminosity may drive a ``superwind'', getting rid of the excess energy in the form of mass loss and not radiation \citep[]{vas93}. In strongly mass-losing situations, the outer boundary becomes dynamical and the hydrostatic assumption breaks down \citep[]{pon21}. Inflated envelopes may be subject to pulsational instabilities, causing strong radial pulsations \citep[]{san15}. All of these effects could influence the stellar structure in the outer parts, and could alter the stellar radius significantly. In some cases the slower parts of the SN ejecta may be captured by the companion, partly obscuring the light or adding to the extinction. Such complications make it difficult to accurately model the radius/temperature evolution of the post-SN companions. 

\begin{figure}
 \centering
 \includegraphics[width=\linewidth]{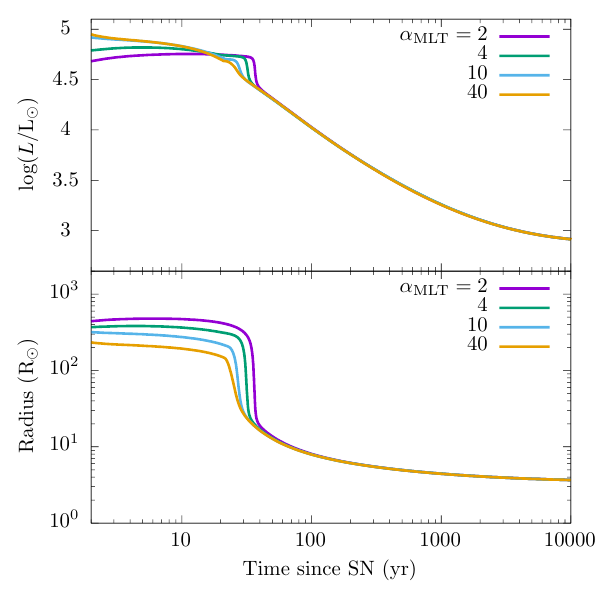}
 \caption{Comparison of the stellar response with different mixing lengths applied for convection. The adopted model parameters were $M_2=5\msun$, $R_2=3.5~\rsun$, $E_\mathrm{exp}=1~B$, $a=13~\rsun$.\label{fig:mlt_comparison}}
\end{figure}

The existence of both the inflated state and the maximum luminosity are strongly correlated. Inflated envelopes are known to emerge as a consequence of inefficient near-surface convection \citep[]{san15}. This is likely due to the way convection is treated in 1D stellar evolution codes. In reality, other potential modes of energy transport may arise such as waves excited by turbulent convection \citep[e.g.][]{mae87} or developing porous layers that reduce the effective opacity \citep[e.g.][]{sha99,beg01,owo04}. To test how extra modes of energy transport may influence our models, we performed the same MESA calculation with different energy transport efficiencies by changing the mixing length parameter $\alpha_\mathrm{MLT}$. In Figure~\ref{fig:mlt_comparison}, we compare the light curve and radius evolution for various mixing lengths. As we increase the mixing length, the maximum achieved luminosity increases while the plateau phase shortens. At the same time, the radius during the inflated phase decreases with larger mixing lengths. However, both the luminosity and radius evolution are identical during the tail phases. We therefore believe that if the main effect of near-Eddington luminosities in stars is to activate more efficient energy transport mechanisms, our bolometric light curve models are still valid especially for the tail phase where the luminosity is sufficiently sub-Eddington. Within our framework, this can be regarded as an uncertainty in $L_\mathrm{max}$ and $\tau_\mathrm{inf}$. On the other hand, if the high luminosity leads to enhanced winds, dynamical pulsations or eruptions, the results may differ in less trivial ways.

The explosion energy for SN2006jc is highly uncertain. Here, we used the observationally inferred value $E_\mathrm{exp}=10~$B \citep[]{tom08}, which is an order of magnitude higher than the typical energies for \acp{sn}Ib. This was calculated under the assumption that the \ac{sn} light curve was powered mostly by radioactive decay. However, it is known that there was some degree of interaction between the \ac{sn} ejecta and dense \ac{csm} from its narrow He~\textsc{i} lines and X-ray emission \citep[]{bro06,imm06,imm08}. Depending on how much the \ac{csm}-interaction contributed to the light curve, the explosion energy estimate could be substantially overestimated. Our constraints on the pre-\ac{sn} separation (Fig~\ref{fig:sep_constraints}) can be much lower depending on what the true explosion energy is.

\subsection{Implications for SESN progenitor formation channels}

The two main channels in creating stripped-envelope stars in binaries are via stable mass transfer and common-envelope evolution. Both channels are expected to be contributing to the whole population of \acp{sesn} \citep[]{pod92}. However, the dominant channel may differ depending on each \ac{sn} subtype, which have differing degrees of remaining hydrogen. Stable mass transfer ceases when the star loses a significant fraction of its envelope and contracts to detach from its Roche lobe. There is still some hydrogen left on top of its core at this point, and the amount depends on the orbital separation and metallicity \citep[]{ouc17,lap20}. Depending on the subsequent wind mass loss and/or later mass transfer episodes, the star may or may not get rid of this remaining hydrogen. Therefore, stable mass transfer channels can create both SNIbc and SNIIb progenitors \citep[]{yoo17}. On the other hand, common-envelope channels are expected to end up with much tighter separations, leaving little to no space for any hydrogen to remain. This implies that common-envelope channels may contribute to the SNIbc population but not much to the SNIIb progenitors. However, a recent model for massive star common-envelopes suggest much wider separations depending on the companion mass, allowing for possible SNIIb progenitor formation\footnote{See also \citet{nai20,shi20} on models involving grazing envelope evolution.} \citep[]{RH22}.

Another distinguishing feature between the two channels is the companion mass. Whether or not the binary undergoes a stable or unstable mass transfer phase depends on many factors such as the evolutionary stage of the donor, mass ratio, the companion star properties, etc \citep[e.g.][]{hje87}. The exact dividing line depends on many uncertain physics such as the response of the star to mass loss \citep[]{sob97,pav15,tem23} and the dynamics of the mass ejected from the system \citep[e.g.][]{mac18}. However, it is qualitatively true that lower mass companions lead to more unstable mass transfer and will experience common-envelope phases. Therefore, post-common-envelope systems will have lower mass companions compared to systems that experienced stable mass transfer. Additionally, stable mass transfer proceeds over a longer timescale, allowing the companion to accrete more material and grow in mass.

The constraints placed on the pre-\ac{sn} binary parameters in the previous section are useful in figuring out which channel formed each \ac{sn} progenitor. For example, the two \acp{sn}IIb SN1993J and SN2011dh are constrained to have relatively massive secondaries on wide orbits (Section~\ref{sec:pre-explosion_image}). Therefore, the progenitor system of SN1993J and SN2011dh likely evolved via stable mass transfer channels. This is in line with various binary evolution modelling attempts for these systems \citep[e.g.][]{pod93,mau04,ben13}. Additionally, the \acp{sn}IIb companion photometry are consistent with being on the main sequence (Fig~\ref{fig:companions}), strengthening the case that these stars were not heated by the \ac{sn} impact. 

For SN2006jc, our model predicts that the companion mass can be as low as $M_2\sim2.5~\msun$, which is too low to sustain a stable mass transfer phase so it likely experienced a common-envelope phase.
If the companion luminosity fades over the next few decades, it will confirm that the companion is a lower mass star, supporting the common-envelope scenario. In our analysis, the pre-SN separation, which is directly related to the post-common-envelope separation, is $\lesssim55~\rsun$ (Figure~\ref{fig:sep_constraints}). The upper edge is somewhat wider than what is typically predicted from the classical energy formalism for common-envelope evolution \citep[e.g.][]{web84}. Therefore, if the companion turns out to be a lower mass star, it will provide strong support for the recent claims that common-envelope phases are much more efficient at ejecting their envelopes in the massive star regime \citep[]{fra19,RH22}.

The other two \acp{sn} on our list do not have additional constraints to help break the degeneracy. However, given that SN2001ig has a very similar companion photometry to SN2011dh, and SN2013ge is similar to SN2006jc (Fig~\ref{fig:companions}), it is tempting to associate them as having similar histories. If this speculation is true, it gives rise to an interesting trend: all the \acp{sn}IIb in this list experienced stable mass transfer and the \acp{sn}Ibc experienced common-envelope phases. It is not appropriate to draw strong conclusions from this small sample size but this trend supports the idea that the distinction between \ac{sn}IIb and \ac{sn}Ibc progenitors is more related to the mode of binary interaction it experienced than other factors such as mass or metallicity \citep[]{fan19,sun23}. Another epoch of companion photometry within the next decade or so will be useful in confirming the true identities of these stars and their evolutionary histories.

\section{Summary and Conclusion}\label{sec:summary}
We present a convenient analytical framework that describes the photometric response of companion stars after being impacted by \acp{sn}. The relatively simple form of the model allows us to solve the inverse problem, i.e. we can infer the pre-\ac{sn} binary properties given the post-\ac{sn} companion observations.

As a demonstration of how our formula can be used to constrain binary interaction physics, we applied our model to five \acp{sesn} that have post-\ac{sn} companion detections (SN1993J, SN2001ig, SN2006jc, SN2011dh, SN2013ge). With one epoch of observations, we are able to place a constraint on the relation between the companion mass and pre-\ac{sn} orbital separation. Combined with other constraints such as the pre-\ac{sn} progenitor photometry and/or post-\ac{sn} companion photometry for multiple epochs, we can determine the evolutionary history of the progenitor system. Based on our analysis, we find that two of the \acp{sn}IIb (SN1993J, SN2011dh) most likely had a fairly massive companion and evolved via stable mass transfer, supporting previous modelling attempts \citep[]{pod93,mau04,ben13}. For the SNIbn 2006jc, we argue that it likely had a lower-mass companion and evolved through a common-envelope phase.

While this framework is useful for rough and quick interpretations from observations, extra work is required to fully understand the response of companion stars after \acp{sn}. Our formula hinges on several key assumptions. First, we assume the amount of stripped mass from the companion is negligible. This is generally not a bad assumption, but can sometimes be large enough to influence the long-term response \citep[]{RH18,che23}. Secondly, we rely on MESA calculations as a basis of our analytical model. Because of the extreme near-Eddington luminosities and non-equilibrium nature of \ac{sn}-heated stars, the intrinsic approximations are inappropriate in some regimes (Section~\ref{sec:caveats}). Further detailed modelling of near-Eddington envelopes is required to predict the accurate response of \ac{sn}-impacted companions.

Continued monitoring of the companions over the next few decades can help pin down the pre-\ac{sn} orbital separations, which is a vital constraint not only to determine their formation channels, but also for understanding binary interaction physics such as the common-envelope phase mechanism and mass transfer efficiency. Such constraints can ultimately help us understand the formation of other objects including X-ray binaries and gravitational wave sources.

In addition to the five \acp{sn} we studied, SN2016gkg may be another interesting source \citep[]{kil22} to continue monitoring. This SNIIb has faded below the pre-SN source but now has excess emission that could indicate a surviving companion. Together with its rich information including the progenitor photometry \citep[]{tar17,kil17} and early light curve \citep[]{arc17,ber18}, we will be able to work out its evolutionary origin.

\section*{Acknowledgements}
The author thanks Alexander Heger and Team COMPAS for useful discussions. The author also thanks the anonymous referee for insightful comments that improved the manuscript. RH is supported by the Australian Research Council (ARC) Centre of Excellence for Gravitational Wave Discovery (OzGrav), through project number CE170100004.

\section*{Data Availability}
The data underlying this article will be shared on reasonable request to the corresponding author.


\bibliographystyle{mnras}

\begin{thebibliography}{}
\makeatletter
\relax
\def\mn@urlcharsother{\let\do\@makeother \do\$\do\&\do\#\do\^\do\_\do\%\do\~}
\def\mn@doi{\begingroup\mn@urlcharsother \@ifnextchar [ {\mn@doi@}
  {\mn@doi@[]}}
\def\mn@doi@[#1]#2{\def\@tempa{#1}\ifx\@tempa\@empty \href
  {http://dx.doi.org/#2} {doi:#2}\else \href {http://dx.doi.org/#2} {#1}\fi
  \endgroup}
\def\mn@eprint#1#2{\mn@eprint@#1:#2::\@nil}
\def\mn@eprint@arXiv#1{\href {http://arxiv.org/abs/#1} {{\tt arXiv:#1}}}
\def\mn@eprint@dblp#1{\href {http://dblp.uni-trier.de/rec/bibtex/#1.xml}
  {dblp:#1}}
\def\mn@eprint@#1:#2:#3:#4\@nil{\def\@tempa {#1}\def\@tempb {#2}\def\@tempc
  {#3}\ifx \@tempc \@empty \let \@tempc \@tempb \let \@tempb \@tempa \fi \ifx
  \@tempb \@empty \def\@tempb {arXiv}\fi \@ifundefined
  {mn@eprint@\@tempb}{\@tempb:\@tempc}{\expandafter \expandafter \csname
  mn@eprint@\@tempb\endcsname \expandafter{\@tempc}}}

\bibitem[\protect\citeauthoryear{{Aldering}, {Humphreys}  \&
  {Richmond}}{{Aldering} et~al.}{1994}]{ald94}
{Aldering} G.,  {Humphreys} R.~M.,   {Richmond} M.,  1994, \mn@doi [\aj]
  {10.1086/116886}, \href
  {https://ui.adsabs.harvard.edu/abs/1994AJ....107..662A} {107, 662}

\bibitem[\protect\citeauthoryear{{Arcavi} et~al.,}{{Arcavi}
  et~al.}{2017}]{arc17}
{Arcavi} I.,  et~al., 2017, \mn@doi [\apjl] {10.3847/2041-8213/aa5be1}, \href
  {https://ui.adsabs.harvard.edu/abs/2017ApJ...837L...2A} {837, L2}

\bibitem[\protect\citeauthoryear{{Begelman}}{{Begelman}}{2001}]{beg01}
{Begelman} M.~C.,  2001, \mn@doi [\apj] {10.1086/320240}, \href
  {https://ui.adsabs.harvard.edu/abs/2001ApJ...551..897B} {551, 897}

\bibitem[\protect\citeauthoryear{{Benvenuto}, {Bersten}  \&
  {Nomoto}}{{Benvenuto} et~al.}{2013}]{ben13}
{Benvenuto} O.~G.,  {Bersten} M.~C.,   {Nomoto} K.,  2013, \mn@doi [\apj]
  {10.1088/0004-637X/762/2/74}, \href
  {https://ui.adsabs.harvard.edu/abs/2013ApJ...762...74B} {762, 74}

\bibitem[\protect\citeauthoryear{{Bersten} et~al.,}{{Bersten}
  et~al.}{2012}]{ber12}
{Bersten} M.~C.,  et~al., 2012, \mn@doi [\apj] {10.1088/0004-637X/757/1/31},
  \href {https://ui.adsabs.harvard.edu/abs/2012ApJ...757...31B} {757, 31}

\bibitem[\protect\citeauthoryear{{Bersten} et~al.,}{{Bersten}
  et~al.}{2018}]{ber18}
{Bersten} M.~C.,  et~al., 2018, \mn@doi [\nat] {10.1038/nature25151}, \href
  {https://ui.adsabs.harvard.edu/abs/2018Natur.554..497B} {554, 497}

\bibitem[\protect\citeauthoryear{{Brown}, {Immler}  \& {Modjaz}}{{Brown}
  et~al.}{2006}]{bro06}
{Brown} P.~J.,  {Immler} S.,   {Modjaz} M.,  2006, The Astronomer's Telegram,
  \href {https://ui.adsabs.harvard.edu/abs/2006ATel..916....1B} {916, 1}

\bibitem[\protect\citeauthoryear{{Chen}, {Rau}  \& {Pan}}{{Chen}
  et~al.}{2023}]{che23}
{Chen} H.-P.,  {Rau} S.-J.,   {Pan} K.-C.,  2023, \mn@doi [\apj]
  {10.3847/1538-4357/acc9af}, \href
  {https://ui.adsabs.harvard.edu/abs/2023ApJ...949..121C} {949, 121}

\bibitem[\protect\citeauthoryear{{Di Stefano}, {Kruckow}, {Gao}, {Neunteufel}
  \& {Kobayashi}}{{Di Stefano} et~al.}{2023}]{dis23}
{Di Stefano} R.,  {Kruckow} M.~U.,  {Gao} Y.,  {Neunteufel} P.~G.,
  {Kobayashi} C.,  2023, \mn@doi [\apj] {10.3847/1538-4357/acae9b}, \href
  {https://ui.adsabs.harvard.edu/abs/2023ApJ...944...87D} {944, 87}

\bibitem[\protect\citeauthoryear{{Drout} et~al.,}{{Drout} et~al.}{2016}]{dro16}
{Drout} M.~R.,  et~al., 2016, \mn@doi [\apj] {10.3847/0004-637X/821/1/57},
  \href {https://ui.adsabs.harvard.edu/abs/2016ApJ...821...57D} {821, 57}

\bibitem[\protect\citeauthoryear{{Eggleton}}{{Eggleton}}{1983}]{egg83}
{Eggleton} P.~P.,  1983, \mn@doi [\apj] {10.1086/160960}, \href
  {https://ui.adsabs.harvard.edu/abs/1983ApJ...268..368E} {268, 368}

\bibitem[\protect\citeauthoryear{{Fang}, {Maeda}, {Kuncarayakti}, {Sun}  \&
  {Gal-Yam}}{{Fang} et~al.}{2019}]{fan19}
{Fang} Q.,  {Maeda} K.,  {Kuncarayakti} H.,  {Sun} F.,   {Gal-Yam} A.,  2019,
  \mn@doi [Nature Astronomy] {10.1038/s41550-019-0710-6}, \href
  {https://ui.adsabs.harvard.edu/abs/2019NatAs...3..434F} {3, 434}

\bibitem[\protect\citeauthoryear{{Fox} et~al.,}{{Fox} et~al.}{2014}]{fox14}
{Fox} O.~D.,  et~al., 2014, \mn@doi [\apj] {10.1088/0004-637X/790/1/17}, \href
  {https://ui.adsabs.harvard.edu/abs/2014ApJ...790...17F} {790, 17}

\bibitem[\protect\citeauthoryear{{Fox} et~al.,}{{Fox} et~al.}{2022}]{fox22}
{Fox} O.~D.,  et~al., 2022, \mn@doi [\apjl] {10.3847/2041-8213/ac5890}, \href
  {https://ui.adsabs.harvard.edu/abs/2022ApJ...929L..15F} {929, L15}

\bibitem[\protect\citeauthoryear{{Fragos}, {Andrews}, {Ramirez-Ruiz}, {Meynet},
  {Kalogera}, {Taam}  \& {Zezas}}{{Fragos} et~al.}{2019}]{fra19}
{Fragos} T.,  {Andrews} J.~J.,  {Ramirez-Ruiz} E.,  {Meynet} G.,  {Kalogera}
  V.,  {Taam} R.~E.,   {Zezas} A.,  2019, \mn@doi [\apjl]
  {10.3847/2041-8213/ab40d1}, \href
  {https://ui.adsabs.harvard.edu/abs/2019ApJ...883L..45F} {883, L45}

\bibitem[\protect\citeauthoryear{{Hirai} \& {Mandel}}{{Hirai} \&
  {Mandel}}{2022}]{RH22}
{Hirai} R.,  {Mandel} I.,  2022, \mn@doi [\apjl] {10.3847/2041-8213/ac9519},
  \href {https://ui.adsabs.harvard.edu/abs/2022ApJ...937L..42H} {937, L42}

\bibitem[\protect\citeauthoryear{{Hirai} \& {Yamada}}{{Hirai} \&
  {Yamada}}{2015}]{RH15}
{Hirai} R.,  {Yamada} S.,  2015, \mn@doi [\apj] {10.1088/0004-637X/805/2/170},
  \href {https://ui.adsabs.harvard.edu/abs/2015ApJ...805..170H} {805, 170}

\bibitem[\protect\citeauthoryear{{Hirai}, {Sawai}  \& {Yamada}}{{Hirai}
  et~al.}{2014}]{RH14}
{Hirai} R.,  {Sawai} H.,   {Yamada} S.,  2014, \mn@doi [\apj]
  {10.1088/0004-637X/792/1/66}, \href
  {https://ui.adsabs.harvard.edu/abs/2014ApJ...792...66H} {792, 66}

\bibitem[\protect\citeauthoryear{{Hirai}, {Podsiadlowski}  \& {Yamada}}{{Hirai}
  et~al.}{2018}]{RH18}
{Hirai} R.,  {Podsiadlowski} {\relax Ph}.,   {Yamada} S.,  2018, \mn@doi [\apj]
  {10.3847/1538-4357/aad6a0}, \href
  {https://ui.adsabs.harvard.edu/abs/2018ApJ...864..119H} {864, 119}

\bibitem[\protect\citeauthoryear{{Hirai}, {Sato}, {Podsiadlowski},
  {Vigna-G{\'o}mez}  \& {Mandel}}{{Hirai} et~al.}{2020}]{RH20}
{Hirai} R.,  {Sato} T.,  {Podsiadlowski} {\relax Ph}.,  {Vigna-G{\'o}mez} A.,
  {Mandel} I.,  2020, \mn@doi [\mnras] {10.1093/mnras/staa2898}, \href
  {https://ui.adsabs.harvard.edu/abs/2020MNRAS.499.1154H} {499, 1154}

\bibitem[\protect\citeauthoryear{{Hjellming} \& {Webbink}}{{Hjellming} \&
  {Webbink}}{1987}]{hje87}
{Hjellming} M.~S.,  {Webbink} R.~F.,  1987, \mn@doi [\apj] {10.1086/165412},
  \href {https://ui.adsabs.harvard.edu/abs/1987ApJ...318..794H} {318, 794}

\bibitem[\protect\citeauthoryear{{Hurley}, {Pols}  \& {Tout}}{{Hurley}
  et~al.}{2000}]{hur00}
{Hurley} J.~R.,  {Pols} O.~R.,   {Tout} C.~A.,  2000, \mn@doi [\mnras]
  {10.1046/j.1365-8711.2000.03426.x}, \href
  {https://ui.adsabs.harvard.edu/abs/2000MNRAS.315..543H} {315, 543}

\bibitem[\protect\citeauthoryear{{Immler}, {Modjaz}  \& {Brown}}{{Immler}
  et~al.}{2006}]{imm06}
{Immler} S.,  {Modjaz} M.,   {Brown} P.~J.,  2006, The Astronomer's Telegram,
  \href {https://ui.adsabs.harvard.edu/abs/2006ATel..934....1I} {934, 1}

\bibitem[\protect\citeauthoryear{{Immler} et~al.,}{{Immler}
  et~al.}{2008}]{imm08}
{Immler} S.,  et~al., 2008, \mn@doi [\apjl] {10.1086/529373}, \href
  {https://ui.adsabs.harvard.edu/abs/2008ApJ...674L..85I} {674, L85}

\bibitem[\protect\citeauthoryear{{Ivanova} et~al.,}{{Ivanova}
  et~al.}{2013}]{iva13}
{Ivanova} N.,  et~al., 2013, \mn@doi [\aapr] {10.1007/s00159-013-0059-2}, \href
  {https://ui.adsabs.harvard.edu/abs/2013A&ARv..21...59I} {21, 59}

\bibitem[\protect\citeauthoryear{{Ivanova}, {Justham}  \& {Ricker}}{{Ivanova}
  et~al.}{2020}]{iva20}
{Ivanova} N.,  {Justham} S.,   {Ricker} P.,  2020, {Common Envelope Evolution},
  \mn@doi{10.1088/2514-3433/abb6f0.
}

\bibitem[\protect\citeauthoryear{{Kilpatrick} et~al.,}{{Kilpatrick}
  et~al.}{2017}]{kil17}
{Kilpatrick} C.~D.,  et~al., 2017, \mn@doi [\mnras] {10.1093/mnras/stw3082},
  \href {https://ui.adsabs.harvard.edu/abs/2017MNRAS.465.4650K} {465, 4650}

\bibitem[\protect\citeauthoryear{{Kilpatrick}, {Coulter}, {Foley}, {Piro},
  {Rest}, {Rojas-Bravo}  \& {Siebert}}{{Kilpatrick} et~al.}{2022}]{kil22}
{Kilpatrick} C.~D.,  {Coulter} D.~A.,  {Foley} R.~J.,  {Piro} A.~L.,  {Rest}
  A.,  {Rojas-Bravo} C.,   {Siebert} M.~R.,  2022, \mn@doi [\apj]
  {10.3847/1538-4357/ac8a4c}, \href
  {https://ui.adsabs.harvard.edu/abs/2022ApJ...936..111K} {936, 111}

\bibitem[\protect\citeauthoryear{{Laplace}, {G{\"o}tberg}, {de Mink}, {Justham}
   \& {Farmer}}{{Laplace} et~al.}{2020}]{lap20}
{Laplace} E.,  {G{\"o}tberg} Y.,  {de Mink} S.~E.,  {Justham} S.,   {Farmer}
  R.,  2020, \mn@doi [\aap] {10.1051/0004-6361/201937300}, \href
  {https://ui.adsabs.harvard.edu/abs/2020A&A...637A...6L} {637, A6}

\bibitem[\protect\citeauthoryear{{Lau}, {Hirai}, {Gonz{\'a}lez-Bol{\'\i}var},
  {Price}, {De Marco}  \& {Mandel}}{{Lau} et~al.}{2022a}]{lau22a}
{Lau} M. Y.~M.,  {Hirai} R.,  {Gonz{\'a}lez-Bol{\'\i}var} M.,  {Price} D.~J.,
  {De Marco} O.,   {Mandel} I.,  2022a, \mn@doi [\mnras]
  {10.1093/mnras/stac049}, \href
  {https://ui.adsabs.harvard.edu/abs/2022MNRAS.512.5462L} {512, 5462}

\bibitem[\protect\citeauthoryear{{Lau}, {Hirai}, {Price}  \& {Mandel}}{{Lau}
  et~al.}{2022b}]{lau22b}
{Lau} M. Y.~M.,  {Hirai} R.,  {Price} D.~J.,   {Mandel} I.,  2022b, \mn@doi
  [\mnras] {10.1093/mnras/stac2490}, \href
  {https://ui.adsabs.harvard.edu/abs/2022MNRAS.516.4669L} {516, 4669}

\bibitem[\protect\citeauthoryear{{Law-Smith} et~al.,}{{Law-Smith}
  et~al.}{2020}]{law20}
{Law-Smith} J. A.~P.,  et~al., 2020, \mn@doi [arXiv e-prints]
  {10.48550/arXiv.2011.06630}, \href
  {https://ui.adsabs.harvard.edu/abs/2020arXiv201106630L} {p. arXiv:2011.06630}

\bibitem[\protect\citeauthoryear{{Liu}, {Tauris}, {R{\"o}pke}, {Moriya},
  {Kruckow}, {Stancliffe}  \& {Izzard}}{{Liu} et~al.}{2015}]{liu15}
{Liu} Z.-W.,  {Tauris} T.~M.,  {R{\"o}pke} F.~K.,  {Moriya} T.~J.,  {Kruckow}
  M.,  {Stancliffe} R.~J.,   {Izzard} R.~G.,  2015, \mn@doi [\aap]
  {10.1051/0004-6361/201526757}, \href
  {https://ui.adsabs.harvard.edu/abs/2015A&A...584A..11L} {584, A11}

\bibitem[\protect\citeauthoryear{{Lyman}, {Bersier}, {James}, {Mazzali},
  {Eldridge}, {Fraser}  \& {Pian}}{{Lyman} et~al.}{2016}]{lym16}
{Lyman} J.~D.,  {Bersier} D.,  {James} P.~A.,  {Mazzali} P.~A.,  {Eldridge}
  J.~J.,  {Fraser} M.,   {Pian} E.,  2016, \mn@doi [\mnras]
  {10.1093/mnras/stv2983}, \href
  {https://ui.adsabs.harvard.edu/abs/2016MNRAS.457..328L} {457, 328}

\bibitem[\protect\citeauthoryear{{MacLeod}, {Ostriker}  \& {Stone}}{{MacLeod}
  et~al.}{2018}]{mac18}
{MacLeod} M.,  {Ostriker} E.~C.,   {Stone} J.~M.,  2018, \mn@doi [\apj]
  {10.3847/1538-4357/aae9eb}, \href
  {https://ui.adsabs.harvard.edu/abs/2018ApJ...868..136M} {868, 136}

\bibitem[\protect\citeauthoryear{{Maeder}}{{Maeder}}{1987}]{mae87}
{Maeder} A.,  1987, \aap, \href
  {https://ui.adsabs.harvard.edu/abs/1987A&A...173..247M} {173, 247}

\bibitem[\protect\citeauthoryear{{Maund}}{{Maund}}{2019}]{mau19}
{Maund} J.~R.,  2019, \mn@doi [\apj] {10.3847/1538-4357/ab2386}, \href
  {https://ui.adsabs.harvard.edu/abs/2019ApJ...883...86M} {883, 86}

\bibitem[\protect\citeauthoryear{{Maund} \& {Smartt}}{{Maund} \&
  {Smartt}}{2009}]{mau09}
{Maund} J.~R.,  {Smartt} S.~J.,  2009, \mn@doi [Science]
  {10.1126/science.1170198}, \href
  {https://ui.adsabs.harvard.edu/abs/2009Sci...324..486M} {324, 486}

\bibitem[\protect\citeauthoryear{{Maund}, {Smartt}, {Kudritzki},
  {Podsiadlowski}  \& {Gilmore}}{{Maund} et~al.}{2004}]{mau04}
{Maund} J.~R.,  {Smartt} S.~J.,  {Kudritzki} R.~P.,  {Podsiadlowski} {\relax
  Ph}.,   {Gilmore} G.~F.,  2004, \mn@doi [\nat] {10.1038/nature02161}, \href
  {https://ui.adsabs.harvard.edu/abs/2004Natur.427..129M} {427, 129}

\bibitem[\protect\citeauthoryear{{Maund} et~al.,}{{Maund} et~al.}{2011}]{mau11}
{Maund} J.~R.,  et~al., 2011, \mn@doi [\apjl]
  {10.1088/2041-8205/739/2/L3710.48550/arXiv.1106.2565}, \href
  {https://ui.adsabs.harvard.edu/abs/2011ApJ...739L..37M} {739, L37}

\bibitem[\protect\citeauthoryear{{Maund}, {Pastorello}, {Mattila}, {Itagaki}
  \& {Boles}}{{Maund} et~al.}{2016}]{mau16}
{Maund} J.~R.,  {Pastorello} A.,  {Mattila} S.,  {Itagaki} K.,   {Boles} T.,
  2016, \mn@doi [\apj] {10.3847/1538-4357/833/2/128}, \href
  {https://ui.adsabs.harvard.edu/abs/2016ApJ...833..128M} {833, 128}

\bibitem[\protect\citeauthoryear{{Moe} \& {Di Stefano}}{{Moe} \& {Di
  Stefano}}{2017}]{moe17}
{Moe} M.,  {Di Stefano} R.,  2017, \mn@doi [\apjs]
  {10.3847/1538-4365/aa6fb610.48550/arXiv.1606.05347}, \href
  {https://ui.adsabs.harvard.edu/abs/2017ApJS..230...15M} {230, 15}

\bibitem[\protect\citeauthoryear{{Moreno}, {Schneider}, {R{\"o}pke}, {Ohlmann},
  {Pakmor}, {Podsiadlowski}  \& {Sand}}{{Moreno} et~al.}{2022}]{mor22}
{Moreno} M.~M.,  {Schneider} F. R.~N.,  {R{\"o}pke} F.~K.,  {Ohlmann} S.~T.,
  {Pakmor} R.,  {Podsiadlowski} {\relax Ph}.,   {Sand} C.,  2022, \mn@doi
  [\aap] {10.1051/0004-6361/202142731}, \href
  {https://ui.adsabs.harvard.edu/abs/2022A&A...667A..72M} {667, A72}

\bibitem[\protect\citeauthoryear{{Naiman}, {Sabach}, {Gilkis}  \&
  {Soker}}{{Naiman} et~al.}{2020}]{nai20}
{Naiman} B.~V.,  {Sabach} E.,  {Gilkis} A.,   {Soker} N.,  2020, \mn@doi
  [\mnras] {10.1093/mnras/stz3224}, \href
  {https://ui.adsabs.harvard.edu/abs/2020MNRAS.491.2736N} {491, 2736}

\bibitem[\protect\citeauthoryear{{Offner}, {Moe}, {Kratter}, {Sadavoy},
  {Jensen}  \& {Tobin}}{{Offner} et~al.}{2022}]{off22}
{Offner} S. S.~R.,  {Moe} M.,  {Kratter} K.~M.,  {Sadavoy} S.~I.,  {Jensen} E.
  L.~N.,   {Tobin} J.~J.,  2022, \mn@doi [arXiv e-prints]
  {10.48550/arXiv.2203.10066}, \href
  {https://ui.adsabs.harvard.edu/abs/2022arXiv220310066O} {p. arXiv:2203.10066}

\bibitem[\protect\citeauthoryear{Ogata, Hirai  \& Hijikawa}{Ogata
  et~al.}{2021a}]{oga21z}
Ogata M.,  Hirai R.,   Hijikawa K.,  2021a, {The observability of inflated
  companion stars after supernovae in massive binaries},
  \mn@doi{10.5281/zenodo.4624586}, \url
  {https://doi.org/10.5281/zenodo.4624586}

\bibitem[\protect\citeauthoryear{{Ogata}, {Hirai}  \& {Hijikawa}}{{Ogata}
  et~al.}{2021b}]{oga21}
{Ogata} M.,  {Hirai} R.,   {Hijikawa} K.,  2021b, \mn@doi [\mnras]
  {10.1093/mnras/stab1439}, \href
  {https://ui.adsabs.harvard.edu/abs/2021MNRAS.505.2485O} {505, 2485}

\bibitem[\protect\citeauthoryear{{Ouchi} \& {Maeda}}{{Ouchi} \&
  {Maeda}}{2017}]{ouc17}
{Ouchi} R.,  {Maeda} K.,  2017, \mn@doi [\apj] {10.3847/1538-4357/aa6ea9},
  \href {https://ui.adsabs.harvard.edu/abs/2017ApJ...840...90O} {840, 90}

\bibitem[\protect\citeauthoryear{{Owocki}, {Gayley}  \& {Shaviv}}{{Owocki}
  et~al.}{2004}]{owo04}
{Owocki} S.~P.,  {Gayley} K.~G.,   {Shaviv} N.~J.,  2004, \mn@doi [\apj]
  {10.1086/424910}, \href
  {https://ui.adsabs.harvard.edu/abs/2004ApJ...616..525O} {616, 525}

\bibitem[\protect\citeauthoryear{{Pavlovskii} \& {Ivanova}}{{Pavlovskii} \&
  {Ivanova}}{2015}]{pav15}
{Pavlovskii} K.,  {Ivanova} N.,  2015, \mn@doi [\mnras] {10.1093/mnras/stv619},
  \href {https://ui.adsabs.harvard.edu/abs/2015MNRAS.449.4415P} {449, 4415}

\bibitem[\protect\citeauthoryear{{Paxton}, {Bildsten}, {Dotter}, {Herwig},
  {Lesaffre}  \& {Timmes}}{{Paxton} et~al.}{2011}]{pax11}
{Paxton} B.,  {Bildsten} L.,  {Dotter} A.,  {Herwig} F.,  {Lesaffre} P.,
  {Timmes} F.,  2011, \mn@doi [\apjs] {10.1088/0067-0049/192/1/3}, \href
  {https://ui.adsabs.harvard.edu/abs/2011ApJS..192....3P} {192, 3}

\bibitem[\protect\citeauthoryear{{Paxton} et~al.,}{{Paxton}
  et~al.}{2013}]{pax13}
{Paxton} B.,  et~al., 2013, \mn@doi [\apjs] {10.1088/0067-0049/208/1/4}, \href
  {https://ui.adsabs.harvard.edu/abs/2013ApJS..208....4P} {208, 4}

\bibitem[\protect\citeauthoryear{{Paxton} et~al.,}{{Paxton}
  et~al.}{2015}]{pax15}
{Paxton} B.,  et~al., 2015, \mn@doi [\apjs] {10.1088/0067-0049/220/1/15}, \href
  {https://ui.adsabs.harvard.edu/abs/2015ApJS..220...15P} {220, 15}

\bibitem[\protect\citeauthoryear{{Paxton} et~al.,}{{Paxton}
  et~al.}{2018}]{pax18}
{Paxton} B.,  et~al., 2018, \mn@doi [\apjs] {10.3847/1538-4365/aaa5a8}, \href
  {https://ui.adsabs.harvard.edu/abs/2018ApJS..234...34P} {234, 34}

\bibitem[\protect\citeauthoryear{{Podsiadlowski}, {Joss}  \&
  {Hsu}}{{Podsiadlowski} et~al.}{1992}]{pod92}
{Podsiadlowski} {\relax Ph}.,  {Joss} P.~C.,   {Hsu} J.~J.~L.,  1992, \mn@doi
  [\apj] {10.1086/171341}, \href
  {https://ui.adsabs.harvard.edu/abs/1992ApJ...391..246P} {391, 246}

\bibitem[\protect\citeauthoryear{{Podsiadlowski}, {Hsu}, {Joss}  \&
  {Ross}}{{Podsiadlowski} et~al.}{1993}]{pod93}
{Podsiadlowski} {\relax Ph}.,  {Hsu} J.~J.~L.,  {Joss} P.~C.,   {Ross} R.~R.,
  1993, \mn@doi [\nat] {10.1038/364509a0}, \href
  {https://ui.adsabs.harvard.edu/abs/1993Natur.364..509P} {364, 509}

\bibitem[\protect\citeauthoryear{{Poniatowski} et~al.,}{{Poniatowski}
  et~al.}{2021}]{pon21}
{Poniatowski} L.~G.,  et~al., 2021, \mn@doi [\aap]
  {10.1051/0004-6361/202039595}, \href
  {https://ui.adsabs.harvard.edu/abs/2021A&A...647A.151P} {647, A151}

\bibitem[\protect\citeauthoryear{{Prentice} et~al.,}{{Prentice}
  et~al.}{2019}]{pre19}
{Prentice} S.~J.,  et~al., 2019, \mn@doi [\mnras] {10.1093/mnras/sty3399},
  \href {https://ui.adsabs.harvard.edu/abs/2019MNRAS.485.1559P} {485, 1559}

\bibitem[\protect\citeauthoryear{{Rimoldi}, {Portegies Zwart}  \&
  {Rossi}}{{Rimoldi} et~al.}{2016}]{rim16}
{Rimoldi} A.,  {Portegies Zwart} S.,   {Rossi} E.~M.,  2016, \mn@doi
  [Computational Astrophysics and Cosmology] {10.1186/s40668-016-0015-4}, \href
  {https://ui.adsabs.harvard.edu/abs/2016ComAC...3....2R} {3, 2}

\bibitem[\protect\citeauthoryear{{Rizzo Smith}, {Kochanek}  \&
  {Neustadt}}{{Rizzo Smith} et~al.}{2023}]{riz23}
{Rizzo Smith} M.,  {Kochanek} C.~S.,   {Neustadt} J.~M.~M.,  2023, \mn@doi
  [\mnras] {10.1093/mnras/stad1483}, \href
  {https://ui.adsabs.harvard.edu/abs/2023MNRAS.523.1474R} {523, 1474}

\bibitem[\protect\citeauthoryear{{Ryder} et~al.,}{{Ryder} et~al.}{2018}]{ryd18}
{Ryder} S.~D.,  et~al., 2018, \mn@doi [\apj] {10.3847/1538-4357/aaaf1e}, \href
  {https://ui.adsabs.harvard.edu/abs/2018ApJ...856...83R} {856, 83}

\bibitem[\protect\citeauthoryear{{Sana} et~al.,}{{Sana} et~al.}{2012}]{san12}
{Sana} H.,  et~al., 2012, \mn@doi [Science]
  {10.1126/science.122334410.48550/arXiv.1207.6397}, \href
  {https://ui.adsabs.harvard.edu/abs/2012Sci...337..444S} {337, 444}

\bibitem[\protect\citeauthoryear{{Sanyal}, {Grassitelli}, {Langer}  \&
  {Bestenlehner}}{{Sanyal} et~al.}{2015}]{san15}
{Sanyal} D.,  {Grassitelli} L.,  {Langer} N.,   {Bestenlehner} J.~M.,  2015,
  \mn@doi [\aap] {10.1051/0004-6361/201525945}, \href
  {https://ui.adsabs.harvard.edu/abs/2015A&A...580A..20S} {580, A20}

\bibitem[\protect\citeauthoryear{{Shaviv}}{{Shaviv}}{1999}]{sha99}
{Shaviv} N.~J.,  1999, \mn@doi [\physrep] {10.1016/S0370-1573(98)00098-2},
  \href {https://ui.adsabs.harvard.edu/abs/1999PhR...311..177S} {311, 177}

\bibitem[\protect\citeauthoryear{{Shishkin} \& {Soker}}{{Shishkin} \&
  {Soker}}{2020}]{shi20}
{Shishkin} D.,  {Soker} N.,  2020, \mn@doi [\mnras] {10.1093/mnras/staa2080},
  \href {https://ui.adsabs.harvard.edu/abs/2020MNRAS.497..855S} {497, 855}

\bibitem[\protect\citeauthoryear{{Soberman}, {Phinney}  \& {van den
  Heuvel}}{{Soberman} et~al.}{1997}]{sob97}
{Soberman} G.~E.,  {Phinney} E.~S.,   {van den Heuvel} E.~P.~J.,  1997, \mn@doi
  [\aap] {10.48550/arXiv.astro-ph/9703016}, \href
  {https://ui.adsabs.harvard.edu/abs/1997A&A...327..620S} {327, 620}

\bibitem[\protect\citeauthoryear{{Sun}, {Maund}, {Hirai}, {Crowther}  \&
  {Podsiadlowski}}{{Sun} et~al.}{2020}]{sun20}
{Sun} N.-C.,  {Maund} J.~R.,  {Hirai} R.,  {Crowther} P.~A.,   {Podsiadlowski}
  {\relax Ph}.,  2020, \mn@doi [\mnras] {10.1093/mnras/stz3431}, \href
  {https://ui.adsabs.harvard.edu/abs/2020MNRAS.491.6000S} {491, 6000}

\bibitem[\protect\citeauthoryear{{Sun}, {Maund}  \& {Crowther}}{{Sun}
  et~al.}{2023}]{sun23}
{Sun} N.-C.,  {Maund} J.~R.,   {Crowther} P.~A.,  2023, \mn@doi [\mnras]
  {10.1093/mnras/stad690}, \href
  {https://ui.adsabs.harvard.edu/abs/2023MNRAS.521.2860S} {521, 2860}

\bibitem[\protect\citeauthoryear{{Taddia} et~al.,}{{Taddia}
  et~al.}{2018}]{tad18}
{Taddia} F.,  et~al., 2018, \mn@doi [\aap] {10.1051/0004-6361/201730844}, \href
  {https://ui.adsabs.harvard.edu/abs/2018A&A...609A.136T} {609, A136}

\bibitem[\protect\citeauthoryear{{Tartaglia} et~al.,}{{Tartaglia}
  et~al.}{2017}]{tar17}
{Tartaglia} L.,  et~al., 2017, \mn@doi [\apjl] {10.3847/2041-8213/aa5c7f},
  \href {https://ui.adsabs.harvard.edu/abs/2017ApJ...836L..12T} {836, L12}

\bibitem[\protect\citeauthoryear{{Temmink}, {Pols}, {Justham}, {Istrate}  \&
  {Toonen}}{{Temmink} et~al.}{2023}]{tem23}
{Temmink} K.~D.,  {Pols} O.~R.,  {Justham} S.,  {Istrate} A.~G.,   {Toonen} S.,
   2023, \mn@doi [\aap] {10.1051/0004-6361/202244137}, \href
  {https://ui.adsabs.harvard.edu/abs/2023A&A...669A..45T} {669, A45}

\bibitem[\protect\citeauthoryear{{Tominaga} et~al.,}{{Tominaga}
  et~al.}{2008}]{tom08}
{Tominaga} N.,  et~al., 2008, \mn@doi [\apj] {10.1086/591782}, \href
  {https://ui.adsabs.harvard.edu/abs/2008ApJ...687.1208T} {687, 1208}

\bibitem[\protect\citeauthoryear{{Van Dyk} et~al.,}{{Van Dyk}
  et~al.}{2011}]{van11}
{Van Dyk} S.~D.,  et~al., 2011, \mn@doi [\apjl]
  {10.1088/2041-8205/741/2/L2810.48550/arXiv.1106.2897}, \href
  {https://ui.adsabs.harvard.edu/abs/2011ApJ...741L..28V} {741, L28}

\bibitem[\protect\citeauthoryear{{Vassiliadis} \& {Wood}}{{Vassiliadis} \&
  {Wood}}{1993}]{vas93}
{Vassiliadis} E.,  {Wood} P.~R.,  1993, \mn@doi [\apj] {10.1086/173033}, \href
  {https://ui.adsabs.harvard.edu/abs/1993ApJ...413..641V} {413, 641}

\bibitem[\protect\citeauthoryear{{Webbink}}{{Webbink}}{1984}]{web84}
{Webbink} R.~F.,  1984, \mn@doi [\apj] {10.1086/161701}, \href
  {https://ui.adsabs.harvard.edu/abs/1984ApJ...277..355W} {277, 355}

\bibitem[\protect\citeauthoryear{{Yoon}, {Dessart}  \& {Clocchiatti}}{{Yoon}
  et~al.}{2017}]{yoo17}
{Yoon} S.-C.,  {Dessart} L.,   {Clocchiatti} A.,  2017, \mn@doi [\apj]
  {10.3847/1538-4357/aa6afe}, \href
  {https://ui.adsabs.harvard.edu/abs/2017ApJ...840...10Y} {840, 10}

\bibitem[\protect\citeauthoryear{{Young}, {Baron}  \& {Branch}}{{Young}
  et~al.}{1995}]{you95}
{Young} T.~R.,  {Baron} E.,   {Branch} D.,  1995, \mn@doi [\apjl]
  {10.1086/309618}, \href
  {https://ui.adsabs.harvard.edu/abs/1995ApJ...449L..51Y} {449, L51}

\bibitem[\protect\citeauthoryear{{Zapartas} et~al.,}{{Zapartas}
  et~al.}{2017}]{zap17}
{Zapartas} E.,  et~al., 2017, \mn@doi [\apj] {10.3847/1538-4357/aa7467}, \href
  {https://ui.adsabs.harvard.edu/abs/2017ApJ...842..125Z} {842, 125}

\bibitem[\protect\citeauthoryear{{Zapartas} et~al.,}{{Zapartas}
  et~al.}{2019}]{zap19}
{Zapartas} E.,  et~al., 2019, \mn@doi [\aap]
  {10.1051/0004-6361/20193585410.48550/arXiv.1907.06687}, \href
  {https://ui.adsabs.harvard.edu/abs/2019A&A...631A...5Z} {631, A5}

\makeatother
\end{thebibliography}



\appendix

\section{Other light curves}
\label{app:otherLC}

\begin{figure}
 \centering
 \includegraphics[width=\linewidth]{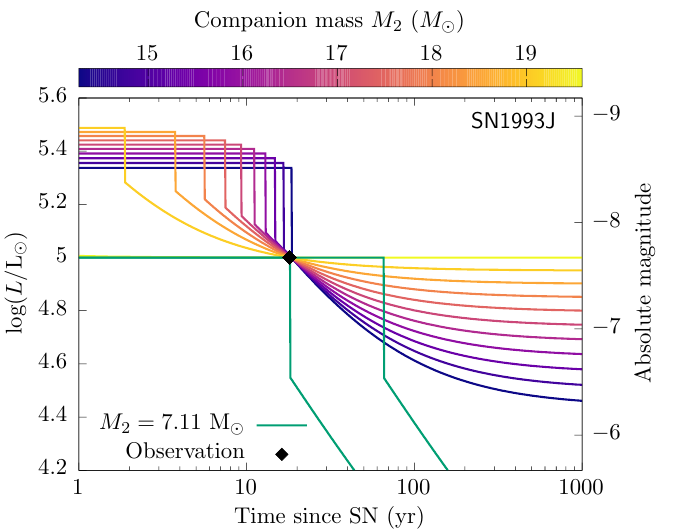}
 \caption{Same as Fig.~\ref{fig:sn2013ge_LC} but for SN1993J.\label{fig:sn1993J_LC}}
\end{figure}

\begin{figure}
 \centering
 \includegraphics[width=\linewidth]{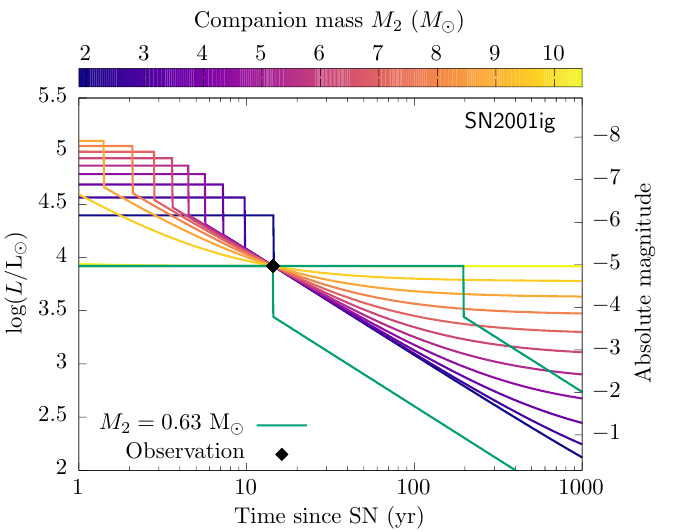}
 \caption{Same as Fig.~\ref{fig:sn2013ge_LC} but for SN2001ig.\label{fig:sn2001ig_LC}}
\end{figure}

\begin{figure}
 \centering
 \includegraphics[width=\linewidth]{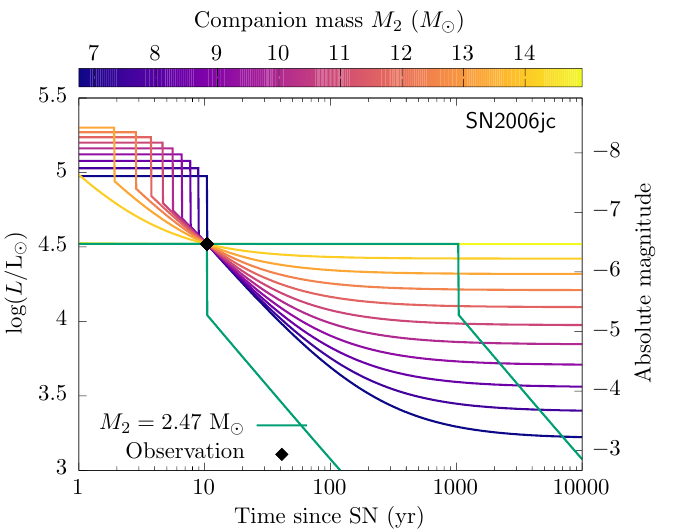}
 \caption{Same as Fig.~\ref{fig:sn2013ge_LC} but for SN2006jc.\label{fig:sn2006jc_LC}}
\end{figure}

\begin{figure}
 \centering
 \includegraphics[width=\linewidth]{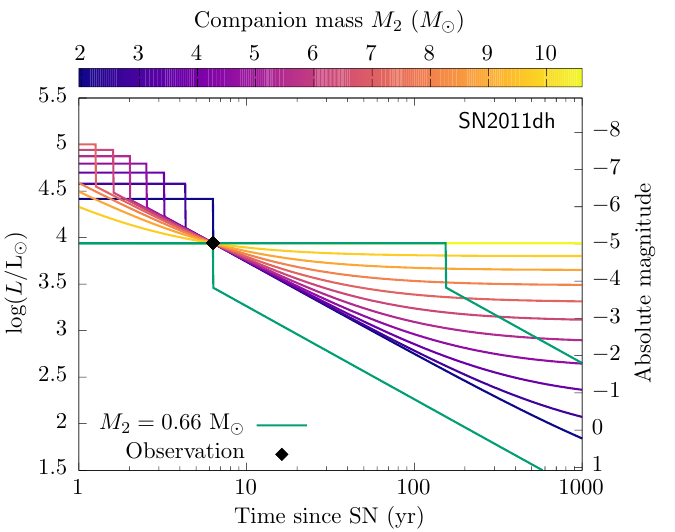}
 \caption{Same as Fig.~\ref{fig:sn2013ge_LC} but for SN2011dh.\label{fig:sn2011dh_LC}}
\end{figure}

Figures~\ref{fig:sn1993J_LC}--\ref{fig:sn2011dh_LC} show the same plots as Figure~\ref{fig:sn2013ge_LC} but for the companion photometry of SN1993J, SN2001ig, SN2006jc and SN2011dh as listed in Table~\ref{tab:photometry}. For the explosion energy, we used the median value within the observationally inferred range (Table~\ref{tab:photometry}). For SN2001ig, the peak of the light curve was not observed so there are no good constraints on the explosion energy. We simply use the canonical value $E_\mathrm{exp}=1~$B for SN2001ig.

\bsp	
\label{lastpage}
\end{document}